\documentclass[12pt]{article}
\usepackage{amssymb}
\usepackage{amsfonts}
\usepackage{amsmath}
\usepackage{latexsym}
\usepackage{bbm}
\usepackage[dvips]{graphicx}
\usepackage{color}
\usepackage{rotating}
\usepackage[colorlinks, citecolor=blue, linkcolor=blue]{hyperref}

\newcommand{\al}{\alpha}

\newcommand{\la}{\lambda}

\newcommand{\eps}{\epsilon}

\renewcommand{\th}{\theta}

\newcommand{\vr}{\!\vec{\,r}}

\newcommand{\sfrac}[2]{{\textstyle\frac{#1}{#2}}}

\newcommand{\pa}{\partial}

\newcommand{\diff}{{\mathrm{d}}}

\newcommand{\beq}{\begin{equation}}
\newcommand{\eeq}{\end{equation}}
\newcommand{\eq}{\end{equation}}
\newcommand{\bea}{\begin{eqnarray}}
\newcommand{\eea}{\end{eqnarray}}

\renewcommand{\and}{{\quad{\rm and}\quad}}
\newcommand{\und}{{\qquad{\rm and}\qquad}}
\renewcommand{\=}{\ =\ }

\advance \textheight by \topskip
\makeatletter
\@addtoreset{equation}{section}

\makeatother

\begin{document}

\begin{titlepage}
\setcounter{page}{0}
\begin{flushright}
ITP--UH--19/15\\
\end{flushright}

\vskip 1.5cm

\begin{center}

{\Huge\bf
On Asteroid Engineering
}

\vspace{12mm}
{\Large Olaf Lechtenfeld 
}
\\[8mm]

\noindent {\em
Institut f\"ur Theoretische Physik and Riemann Center for Geometry and Physics\\
Leibniz Universit\"at Hannover \\
Appelstra\ss{}e 2, 30167 Hannover, Germany }\\[6pt]
{Email: olaf.lechtenfeld@itp.uni-hannover.de}

\vspace{20mm}

\begin{abstract}
\noindent
I pose the question of maximal Newtonian surface gravity on a homogeneous body
of a given mass and volume but with variable shape. In other words, given an
amount of malleable material of uniform density, how should one shape it in
order for a microscopic creature on its surface to experience the largest 
possible weight? After evaluating the weight on an arbitrary cylinder,
at the axis and at the equator and comparing it to that on a spherical ball,
I solve the variational problem to obtain the shape which optimizes the surface
gravity in some location. The boundary curve of the corresponding solid of
revolution is given by \ $(x^2+z^2)^3-(4\,z)^2=0$ \ or \ $r(\th)=2\sqrt{\cos\th}$, 
and the maximal weight (at $x=z=0$) exceeds that on a solid sphere by a factor 
of $\sfrac35\root3\of5$, which is an increment of $2.6\%$. Finally, the values
and the achievable maxima are computed for three other families of shapes. 
\end{abstract}

\vspace{30mm}

{\em in memory of Pascal Gharemani, 03/08/1953 -- 03/08/2015}

\end{center}

\end{titlepage}

\section{Introduction}

\noindent
In the spring of 1996 I was visiting the City College of New York for a month,
in order to pursue a research project with Stuart Samuel, who was a professor
at City University of New York at the time, and to run in the 100th Boston marathon. 
Several evenings and part of weekends I'd spend with our mutual friend Pascal Gharemani, 
a tennis coach and instructor at Trinity School (a private high school on West 
91st Street in Manhattan). Typically we would go dining, visit places or fly kites.
Pascal had an Iranian background but grew up in Versailles near Paris before
moving to the US. My wife and I had come to know him during my postdoc years
at City College (1987--90), when we would meet weekly at various restaurants in the
Columbia University neighborhood for an evening of French conversation.
He was important for our socialization in Manhattan and had grown into a good friend.
Pascal was a very curious individual, with a great sense of humor and always ready 
to engage in discussions about savoir vivre, philosophy, and the natural sciences.
Regarding the latter, he regularly pondered phenomena and questions which involved physics.
Lacking a formal science training, he would go to great lengths and try his physicist 
friends for explanations. %until satisfied. 

So one evening in 1996 he shared his musings about the gravitational force of a long 
and homogeneous rod, as it is felt by a (say, minuscule) creature crawling on its surface. 
Clearly, the mass points in its neighborhood are mainly responsible for creating the force.
On one hand, at the end of the rod, the nearby mass is fewer than elsewhere, but it is all 
pulling roughly in the same direction. On the other hand, around the middle part of the rod, 
twice as much mass points are located near the creature, yet their gravitational forces point 
to almost opposing directions and hence tend to cancel each other out. So which location 
gives more weight to the mini-bug? Where along the rod is its surface gravity largest? 

This was a typical `Pascal question', and my immediate response was: ``That's an easy one.
Let me just compute it.'' Well, easier said then done. For the mid-rod position 
the resulting integrals were too tough to perform on the back of an envelope.
To simplify my life, I persuaded Pascal to modify the problem. 
Let us vary not the position of the bug but the geometry of its planet: 
keep the bug sitting on the top of a cylinder, and compare a long rod with 
a slim disk of the same volume and mass. Then it was not too hard to calculate the surface 
gravity as a function of the ratio of the cylinder's diameter to its length.
To our surprise, in a narrow window of this parameter the weight of the bug exceeds the
value for a spherical ball made from the same material. This finding inspired us to
generalize the question to another level: Given a bunch of homogeneous material
(fixed volume and density, hence total mass), for which shape is the gravitational
force somewhere on its surface maximized? Thus, the idea of ``asteroid engineering'' was born.

After solving the problem and comparing the result with a few other geometries, 
I put the calculations aside and forgot about them. Four years later, when teaching 
Mathematical Methods for physics freshmen, I was looking for a good student exercise
in variational calculus. Coming across my notes from 1996, I realized they can be turned
into an unorthodox, charming and slightly challenging homework problem. 
And so I did, posing the challenge in the summer of 2000~\cite{exercise1} and again 
in 2009~\cite{exercise2}, admittedly with mixed success.\footnote{
In 2002, the problem also occurred in a physics quizz page~\cite{kantor} and later 
in the textbook~\cite{morin}.}
But let the reader decide!

\newpage{}

\begin{figure}[!ht]
\centerline{%\lower8pt\hbox{
\includegraphics[width=8cm]{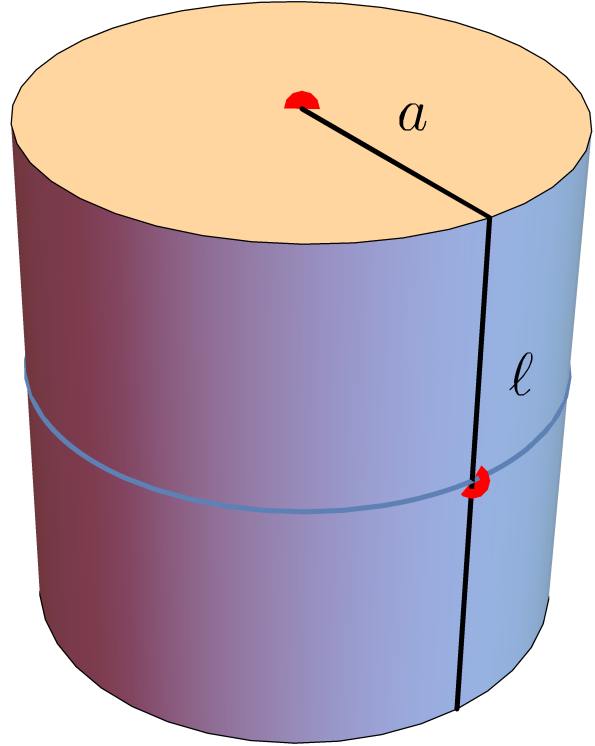}}%}
\caption{Geometry of massive cylinder}
\label{fig:1}
\end{figure}
\vspace{1cm}

\section{Surface gravity of a homogeneous massive cylinder}

\noindent
It is textbook material how to compute the Newtonian gravitational field~$\vec G(\vr)$
generated by a given three-dimensional static mass distribution~$\rho(\vr')$. 
In the absence of symmetry arguments, it involves a three-dimensional integral collecting 
the contributions
\begin{equation}
\diff\vec G(\vr,\vr') \= 
\gamma\,\rho(\vr')\,\frac{\vr'\!-\vr}{|\vr'\!-\vr|^3}\,\diff^3\vr'
\end{equation}
produced by the masses at positions $\vr'$, with $\gamma$ denoting the gravitational constant.
For the case of a solid homogeneous body~$B$ of volume~$V$ and total mass~$M$, clearly
$\rho(\vr')=M/V$ is constant, and one gets
\begin{equation} \label{surfacegravity}
\vec G(\vec r) \= \gamma\,\sfrac{M}{V} \int_B\!\diff^3\vr' \frac{\vec{e}_{\vr'\!-\vr}}{(\vr'\!-\vr)^2}\ ,
\end{equation} 
where $\vec{e}_{\vr'\!-\vr}$ is the unit vector pointing from the observer (at $\vr$) to the mass point
at~$\vr'$. The surface gravity (specific weight of a probe) located somewhere on the surface
$\partial B$ of my solid is obtained by simply restricting $\vr$ to $\partial B$.

One might think of simplifying the task by computing the gravitational potential rather than the field,
since the corresponding integral is scalar and appears to be easier. However, evaluating the surface
gravity then requires taking a gradient in the end and thus keeping at least an infinitesimal dependence 
on a coordinate normal to the surface. Retaining this additional parameter until finally computing the
derivative of the potential with respect to it before setting it to zero yields no calculational gain
over a direct computation of~$\vec G$.

The original question of Pascal concerned a cylindrical rod, whose length and radius I denote 
by $\ell$ and $a$, respectively, so that $V = \pi a^2 \ell$. The integral above has dimension of length,
and I shall scale out a factor of~$\ell$ to pass to dimensionless quantities. For the remaining 
dimensionless parameter I choose the ratio of diameter to length of the cylinder, $t:=2a/\ell$, see Fig.~1.
I shall frequently have to express some of the four quantities $a$, $\ell$, $t$ and $V$ in terms of
a pair of the others, so let me display the complete table of the relations,
\begin{equation}
\begin{aligned}
a &\= \ell\,t/2 \= \sqrt{V/(\pi\ell)} \= \root3\of{V\,t/(2\pi)} \\[4pt]
\ell &\= 2a/t \= V/(\pi a^2) \ \,\= \root3\of{4 V/(\pi t^2)} \\[4pt]
t &\= 2a/\ell \= 2\pi a^3/V \ \;\= \sqrt{4\,V/(\pi\ell^3)} \\[6pt]
V &\= \pi a^2\ell \= 2\pi a^3/t \ \ \ \= \pi\ell^3 t^2/4\ .
\end{aligned}
\end{equation}
Pascal's problem was to compare for this cylinder the surface gravity at the symmetry axis point 
to the one at a point on the mid-circumference or equator. Let me treat both cases in turn.

\subsection{Surface gravity at the axis}

\noindent
Naturally I employ cylindrical coordinates $(z,\rho,\phi)$ for $\vr'$ and put the symmetry axis point 
in the origin. With $\vr=0$ the expression~(\ref{surfacegravity}) then becomes
\begin{equation}
\begin{aligned}
\vec G(0) &\= \gamma\,\sfrac{M}{V} \int_0^\ell\!\diff{z}\int_0^a\!\diff{\rho}\,\rho\int_0^{2\pi}\!\!\!\diff\phi\
(z^2+\rho^2)^{-3/2}\ \Bigl( \begin{smallmatrix} \rho\cos\phi \\ \rho\sin\phi \\ -z \end{smallmatrix} \Bigr)
\\[4pt]
&\= -2\pi\,\gamma\,\sfrac{M}{V} \int_0^\ell\!\diff{z}\int_0^a\!\diff{\rho}\
\frac{\rho\,z}{(z^2+\rho^2)^{3/2}}\ \vec{e}_z
\ =:\ -G_a\,\vec{e}_z\ .
\end{aligned}
\end{equation}
The $\rho$ and $z$ integrals are elementary,
\begin{equation}
\begin{aligned}
G_a 
&\= 2\pi\gamma\,\sfrac{M}{V} \int_0^\ell\!\diff{z}\int_0^a\!\diff{\rho}\ \frac{\rho\,z}{(z^2+\rho^2)^{3/2}} 
 \= 2\pi\gamma\,\sfrac{M}{V} \int_0^\ell\!\diff{z}\ \Bigl[ \frac{z}{\sqrt{z^2+\rho^2}} \Bigr]_0^a  \\[4pt]
&\= 2\pi\gamma\,\sfrac{M}{V} \int_0^\ell\!\diff{z}\ \Bigl\{ 1 - \frac{z}{\sqrt{z^2+a^2}} \Bigr\}
 \= 2\pi\gamma\,\sfrac{M}{V} \,\Bigl[ z - \sqrt{z^2+a^2} \Bigr]_0^\ell \\[4pt]
&\= 2\pi\gamma\,\sfrac{M}{V} \,\Bigl\{ \ell + a - \sqrt{\ell^2+a^2} \Bigr\}
 \= 2\pi\gamma\,\sfrac{M}{V}\,\ell\,\Bigl\{ 1 + \sfrac{t}{2} - \sqrt{1+\sfrac{t^2}{4}} \Bigr\}\ .
\end{aligned}
\end{equation}
It is a bit curious that the result is symmetric under the exchange of $\ell$ and~$a$,
and so in the thin rod ($a\to0$) and thin disk ($\ell\to0$) limits one finds that
\begin{equation}
G_a \= 2\pi\gamma\,\sfrac{M}{V}\,a\,\bigl\{ 1 - \sfrac{a}{2\ell} + \ldots \bigr\} \und
G_a \= 2\pi\gamma\,\sfrac{M}{V}\,\ell\,\bigl\{ 1 - \sfrac{\ell}{2a} + \ldots \bigr\}\ ,
\end{equation}
respectively, with $a^2\ell=V/\pi$ fixed of course.

Apart from the linear dependence on the gravitational constant~$\gamma$ 
and the mass density~$\sfrac{M}{V}$, the surface gravity must carry a dimensional length factor,
which choose to be the cylinder length~$\ell$. 
However, $\ell$, $t$ and $V$ are obviously related, and for comparing different shapes 
of the same mass and volume it is preferable to eliminate $\ell$ in favor of $V$ and~$t$. 
The resulting expression for the surface gravity has the universal form
\begin{equation}
G \= \textrm{(numerical factor)}\ \times\ \gamma\,M\,V^{-2/3}\ \textrm{(shape function)}\ , 
\end{equation}
where the shape function depends on dimensionless parameters like~$t$ only.
For the case at hand, I obtain
\begin{equation}
G_a \= 2^{5/3} \pi^{2/3}\,\gamma\,M\,V^{-2/3}\,t^{-2/3}\,
\Bigl\{ 1 + \sfrac{t}{2} - \sqrt{1+\sfrac{t^2}{4}} \Bigr\}\ .
\end{equation}
The asymptotic behavior for a thin rod ($t\to0$) and for a thin disk ($t\to\infty$) takes the form
\begin{equation}
G_a \= 2^{5/3} \pi^{2/3}\,\gamma\,M\,V^{-2/3}\,\times\, \begin{cases} 
\sfrac{1}{2}t ^{1/3}- \sfrac{1}{8}t^{4/3} + \sfrac{1}{128}t^{10/3} + O(t^{16/3}) 
& \textrm{for} \quad t\to 0 \\[8pt]
t^{-2/3} - t^{-5/3} + t^{-11/3} + O(t^{-17/3}) 
& \textrm{for} \quad t\to\infty \end{cases}\ .
\end{equation}

\subsection{Surface gravity at the equator}

\noindent
This is the harder case, as it lacks the cylindrical symmetry. 
Naturally putting the origin of the cylindrical coordinate system at the
cylinder's center of mass, hence $\vr=(a,0,0)^\top$, the surface gravity integral~(\ref{surfacegravity}) reads
\begin{equation}
\begin{aligned}
\vec G(a) 
&\= \gamma\,\sfrac{M}{V} \int_{-\ell/2}^{\ell/2}\!\!\!\diff{z}\int_0^a\!\diff{\rho}\,\rho\int_0^{2\pi}\!\!\!\diff\phi\
\bigl([\rho\cos\phi-a]^2+[\rho\sin\phi]^2+z^2\bigr)^{-3/2}\ 
\Bigl( \begin{smallmatrix} \rho\cos\phi-a \\ \rho\sin\phi \\ z \end{smallmatrix} \Bigr) \\[4pt]
&\= \gamma\,\sfrac{M}{V} \int_{-\ell/2}^{\ell/2}\!\!\!\diff{z}\int_0^a\!\diff{\rho}\,\rho\int_0^{2\pi}\!\!\!\diff\phi\
\frac{\rho\cos\phi-a}{(z^2+a^2+\rho^2-2\,a\rho\cos\phi)^{3/2}}\ \vec{e}_x \\[4pt]
&\= 2\,\gamma\,\sfrac{M}{V}\,\ell\, \int_0^{1/2}\!\!\diff{u}\int_0^1\!\diff{v}\int_0^{2\pi}\!\!\!\diff\phi\
\frac{v\,(v\cos\phi-1)}{u^2\ell^2/a^2+1+v^2-2\,v\cos\phi)^{3/2}}\ \vec{e}_x 
\ =:\ -G_m\,\vec{e}_x\ ,
\end{aligned}
\end{equation}
where I employed the $z\leftrightarrow-z$ symmetry and substituted $z=u\,\ell$ and $\rho=v\,a$ 
for a dimensionless integral.
The $u$ integration is elementary,
\begin{equation} \label{hardintegral}
\begin{aligned} 
G_m
&\= \gamma\,\sfrac{M}{V}\,\ell\,\int_0^1\!\diff{v}\int_0^{2\pi}\!\!\!\diff\phi\
\frac{v\,(1-v\cos\phi)/(1+v^2-2\,v\cos\phi)}{\sqrt{\ell^2/(4a^2)+1+v^2-2\,v\cos\phi}} \\[4pt]
&\= 2\,\gamma\,\sfrac{M}{V}\,\ell\,\int_0^1\!\diff{v}\int_{-1}^{1}\!\!\diff{w}\
\frac{v\,(1-v\,w)/(1+v^2-2\,v\,w)}{\sqrt{(1-w^2)(t^{-2}+1+v^2-2\,v\,w)}}\ ,
\end{aligned}
\end{equation}
after substituting $\cos\phi=w$ and using the definition $2a/\ell=t$.

The remaining double integrals leads to lengthy expressions
in terms of complete elliptic integrals, which I do not display here.
For $t\to\infty$ it diverges logarithmically.
It is possible, however, to extract the limiting behavior for $t\to0$ as
\begin{equation}
G_m \= 2\pi\,\gamma\,\sfrac{M}{V}\,a\,\bigl\{ 1 - O(\sfrac{a}{\ell}) \bigr\}\ ,
\end{equation}
which in leading order surprisingly agrees with that of $G_a$.

\newpage{}

\begin{figure}[!h]
%\centerline{%\lower12pt\hbox{
\includegraphics[width=16cm]{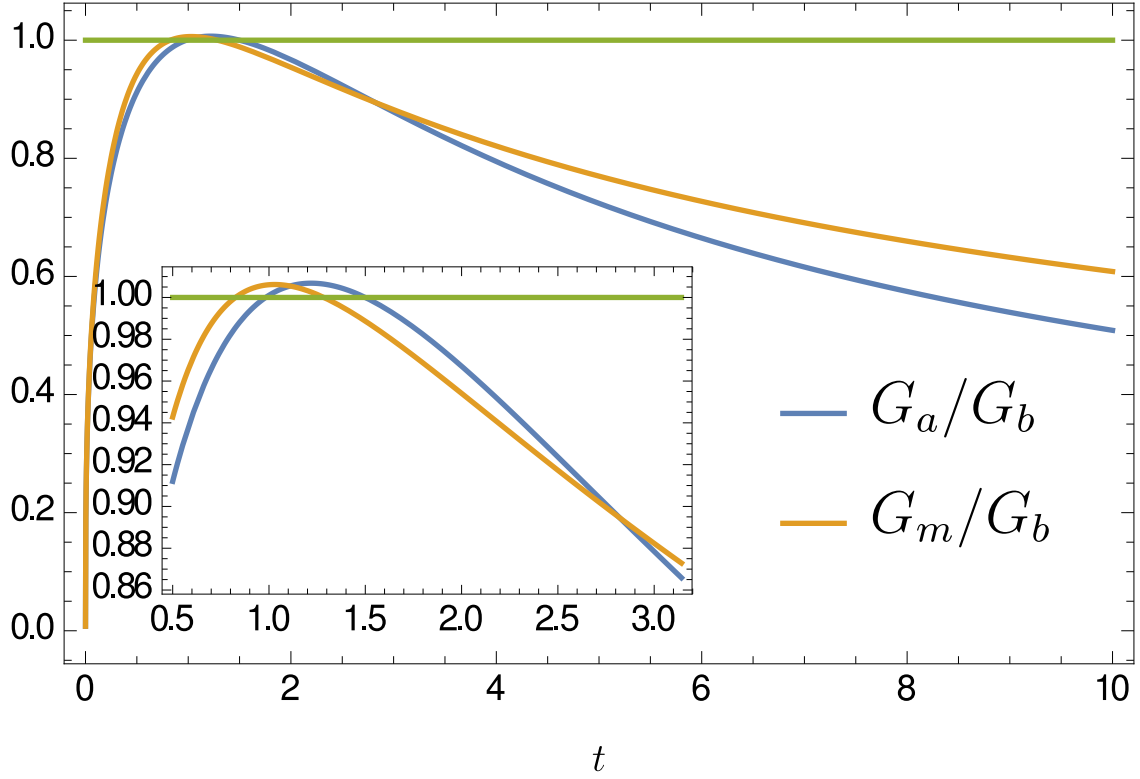}%}%}
\caption{Cylinder surface gravity on symmetry axis and mid-circumference}
\label{fig:2}
\end{figure}
\vspace{0.5cm}

\subsection{Comparison with a spherical ball}

\noindent
To get a feeling for these results, it is natural to compare them with the surface gravity of a homogeneous ball
of the same mass~$M$ and density, thus of radius
\begin{equation}
r_b \= \bigl(\sfrac{4\pi}{3}\bigr)^{-1/3} V^{1/3}\ .
\end{equation}
The surface gravity $\vec G(r_b)=-G_b\,\vec{e}_r$ of the latter is well known,
\begin{equation}
G_b \= \gamma\,M/r_b^2 \= \gamma\,\sfrac{M}{V}\,\sfrac{4\pi}{3}\,r_b \= 
\bigl(\sfrac{4\pi}{3}\bigr)^{2/3} \gamma\,M\,V^{-2/3}\ .
\end{equation}
Hence, the relation of the cylindrical to the spherical surface gravity is
\begin{equation}
\frac{G_a}{G_b} \= 2\pi\,\bigl(\sfrac{\pi}{4}\bigr)^{-1/3} t^{-2/3}\,
\Bigl\{ 1 + \sfrac{t}{2} - \sqrt{1+\sfrac{t^2}{4}} \Bigr\} \Big/ \bigl(\sfrac{4\pi}{3}\bigr)^{2/3}
\= \root 3 \of {18}\;t^{-2/3}\,\Bigl\{ 1 + \sfrac{t}{2} - \sqrt{1+\sfrac{t^2}{4}} \Bigr\}\ ,
\end{equation}
for the axis position, see~Fig.~2.
Surprisingly, in the interval
\begin{equation}
t\ \in\ \bigl[ \sfrac49(2\sqrt{13}{-}5)\ ,\ \sfrac32 \bigr] \ \approx\ 
\bigl[ 0.98271\ ,\ 1.50000 \bigr]
\end{equation}
the weight on the cyclinder's axis exceeds that on the reference ball!
Indeed, its maximal value is attained at
\begin{equation}
t_a \= \sfrac14(9-\sqrt{17}) \ \approx\ 1.21922 \qquad\Rightarrow\qquad
\sfrac{G_a}{G_b}\big|_{\textrm{max}} \= \sfrac{G_a}{G_b}(t_a) \ \approx\ 1.00682\ .
\end{equation}
The asymptotic behavior is easily deduced to be
\begin{equation}
\frac{G_a}{G_b} \= \root3\of{\sfrac{9\pi}{2}}\,\frac{a}{V^{1/3}}\,\Bigl(1-O\bigl(\sfrac{a^3}{V}\bigr)\Bigr) \und
\frac{G_a}{G_b} \= \root3\of{\sfrac{9\pi}{2}}\,\frac{\ell}{V^{1/3}}\,\Bigl(1-O\bigl(\sqrt{\sfrac{\ell^3}{V}}\bigr)\Bigr)\ ,
\end{equation}
for $a\to0$ and $\ell\to0$, respectively.

For the equatorial position's surface gravity I do not have an analytic expression, 
only its limiting forms
\begin{equation}
\frac{G_m}{G_b} \ \sim\ \root3\of{\sfrac{9\pi}{2}}\,\frac{a}{V^{1/3}} \ \approx\  
2.41799\,\frac{a}{V^{1/3}} \und
\frac{G_m}{G_b} \ \sim\ 0.36813\,\frac{\ell}{V^{1/3}}\Bigl|\log\frac{\ell}{V^{1/3}}\Bigr| 
\end{equation}
for $a\to0$ and $\ell\to0$, respectively.
Numerical analysis shows that $G_m/G_b$ (see Fig.~2) attains a maximum at
\begin{equation}
t_m \ \approx \ 1.02928 \qquad\Rightarrow\qquad
\sfrac{G_m}{G_b}\big|_{\textrm{max}} \ \approx \ 1.00619\ .
\end{equation}
Furthermore, for any given shape in an asymptotic regime, the equatorial position is superior to the axis one.
Only in the interval $1.10948 \lesssim t \lesssim 2.82154$ is our mini-bug heavier on the axis.

\section{Which shape maximizes the surface gravity?}

\noindent
This finding suggests the question: Can one do better than the cylinder with a clever choice of shape?
It turns the problem into a variational one. Suppose I have by some means discovered the homogeneous 
body~$\bar B$ which, for fixed mass and volume, yields the maximally possible gravitational pull in some 
location on its surface. 
Without loss of generality I can put this point to the origin of my coordinate system and orient the solid
in such a way that its outward normal in this point aims in the positive $z$~direction,
so gravity pulls downwards as is customary.
Expressing the surface gravity at this position for an arbitrary body~$B$ as a functional of its shape,
then~$\bar B$ must maximize this functional, under the constraint of fixed mass and volume.
The following three features of the optimal shape are evident:
\begin{itemize}
\addtolength{\itemsep}{-8pt}
\item It does not have any holes, so has just a single boundary component
\item It is convex
\item It is rotationally symmetric about the normal at the origin
\end{itemize}
\vspace{-1cm}
\begin{figure}[!ht]
\begin{minipage}[t]{0.5\textwidth}
\begin{picture}(140,260)(60,65)
\put(0,0){\includegraphics[width=12cm,trim=0 0 0 70,clip]{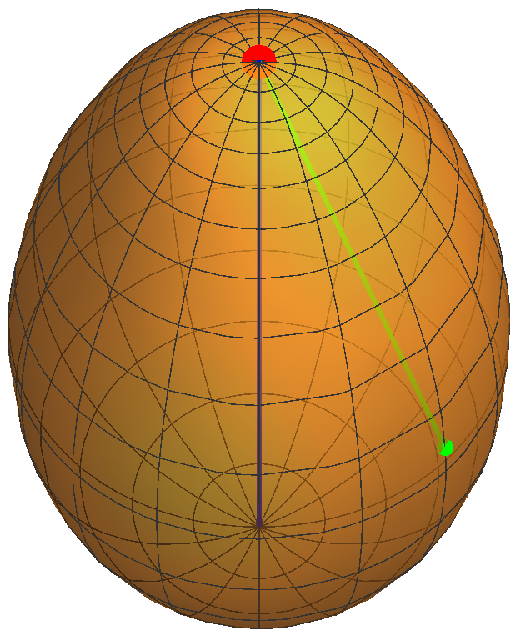}}
\end{picture}
\end{minipage}
\begin{minipage}[t]{0.5\textwidth}
\begin{picture}(140,180)(-50,0)
\put(0,0){\includegraphics[width=6cm]{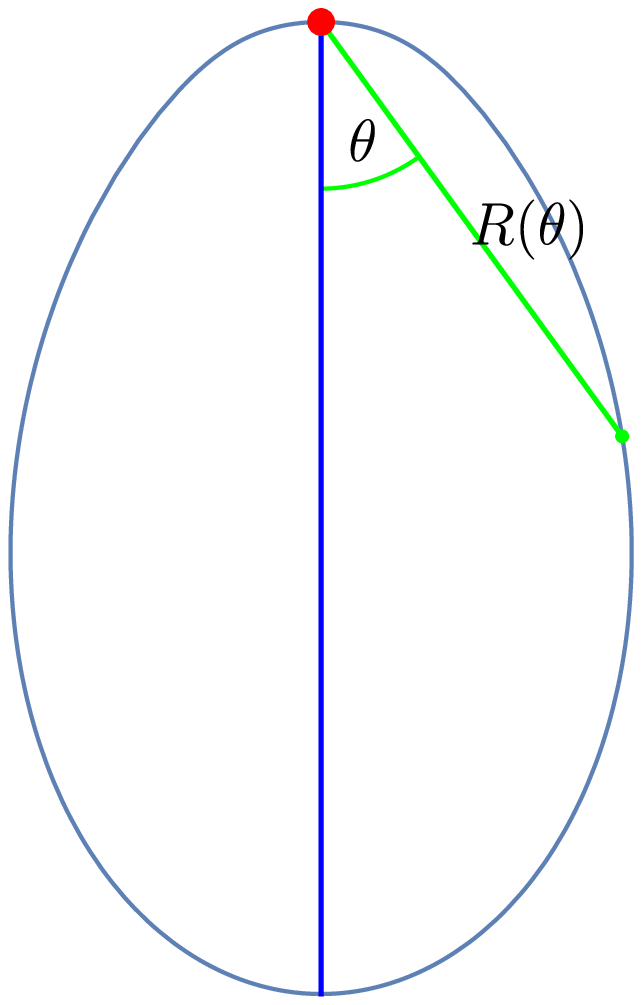}}
\end{picture}
\end{minipage}
\caption{ Parametrization of surface of revolution $\partial B$}
\label{fig:3}
\end{figure}
\vspace{1cm}
These facts imply that the surface~$\partial B$ may be parametrized as in Fig.~3,
\begin{equation}
\partial B \= \bigl\{ R(\th) (\sin\th\cos\phi,\sin\th\sin\phi,-\cos\th)^\top \ \big| \
0\le\th\le\sfrac{\pi}{2}\;,\ 0\le\phi<2\pi \bigr\} \ ,
\end{equation}
with $R(\th)\ge0$ and $R(\sfrac{\pi}{2})=0$.
The function~$R(\th)$ (which may be extended via $R(-\th)=R(\th)$) completely describes 
the shape of the solid of revolution~$B$. 
It may be viewed as the boundary curve of the intersection of~$B$ with the $xz$~plane.
Its convexity implies the condition
\begin{equation}
\bigl(\sfrac1{R(\th)}\bigr)''+\sfrac1{R(\th)}\ \ge\ 0\ .
\end{equation}
Employing the symmetry under reflection on the rotational axis,
\begin{equation}
S\,:\ (\th,\phi)\ \mapsto\ (\th,\phi{+}\pi)\ \sim\ (-\th,\phi)\ ,
\end{equation}
the surface gravity functional (\ref{surfacegravity}) then reads
\begin{equation} 
\vec{G}[R] \= \gamma\,\sfrac{M}{V} \int_B\!\frac{\diff^3\vr}{r^2}\ \sfrac12(\vec{e}_{\vr}+S\vec{e}_{\vr})
\ =:\ -G[R]\,\vec{e}_z\ ,
\end{equation}
\begin{equation} \label{functional}
G[R]\= 2\pi\,\gamma\,\sfrac{M}{V} \int_0^1\!\diff\cos\th \int_0^{R(\th)}\!\!\diff{r}\;\cos\th 
\= 2\pi\,\gamma\,\sfrac{M}{V} \int_0^1\!\diff\cos\th\ R(\th) \cos\th\ .
\end{equation}

It is to be maximized with the mass (and thus the volume) kept fixed,
\begin{equation} \label{constraint}
M[R] \= \sfrac{M}{V} \int_B\!\diff^3\vr 
\= 2\pi\,\sfrac{M}{V} \int_0^1\!\diff\cos\th \int_0^{R(\th)}\!\!r^2\diff{r}
\= \sfrac{2\pi}{3}\,\sfrac{M}{V} \int_0^1\!\diff\cos\th\ R(\th)^3
\ \buildrel ! \over = \ M\ .
\end{equation}
Such constrained variations are best treated by the method of Lagrange multipliers,
which here instructs me to combine the two functionals to
\begin{equation}
2\pi\,\sfrac{M}{V}\,U[R,\la] \= G[R]\ -\ \la\bigl(M[R]-M\bigr)\ ,
\end{equation}
introducing a Lagrange multiplier~$\la$  (a real parameter to be fixed subsequently).
More explicitly,
\begin{equation}
U[R,\la] \= \int_0^1\!\diff\cos\th\ \bigl[ \gamma\,R(\th)\cos\th\ -\ \sfrac13\la\,R(\th)^3 \bigr]
\ -\ \la\,\sfrac{V}{2\pi}\ ,
\end{equation}
so $\pa_\la U=0$ clearly fixes the volume of $B$ to be equal to $V$.
Demanding that, for $\la$ fixed but arbitrary, $U$ is stationary under any variation of the boundary curve, 
$R\mapsto R+\delta R$, determines $R=R_\la$:
\begin{equation}
0 \= \delta U[R_\la,\la] 
\= \int_0^1\!\diff\cos\th\ \delta R(\th)\ \bigl[ \gamma\,\cos\th\ -\ \la\,R_\la(\th)^2 \bigr]\ ,
\end{equation}
so I immediately read off
\begin{equation}
R_\la(\th) \= \sqrt{\sfrac{\gamma}{\la}\,\cos\th}\ .
\end{equation}

It remains to compute the value $\bar\la$ of the Lagrange multiplier by inserting the solution $R_\la$ into
the constraint~(\ref{constraint}),
\begin{equation}
M\ \buildrel ! \over = \ M[R_{\bar\la}] 
\= \sfrac{2\pi}{3}\,\sfrac{M}{V} \int_0^1\!\diff\cos\th\ \bigl( \sfrac{\gamma}{\bar\la}\,\cos\th \bigr)^{3/2}
\= \sfrac{4\pi}{15}\,\sfrac{M}{V}\,\bigl(\sfrac{\gamma}{\bar\la}\bigr)^{3/2}\ ,
\end{equation}
yielding \ $\bar\la=\bigl(\sfrac{4\pi}{15\,V}\bigr)^{2/3}\gamma$ \ and hence the complete solution
as displayed in Fig.~4,
\begin{equation} \label{solution}
\bar{R}(\th)\ :=\ R_{\bar\la}(\th) \= 2\,R_0\,\sqrt{\cos\th} 
\qquad\textrm{with}\qquad (2\,R_0)^3 \= \sfrac{15}{4\pi}\,V \ .
\end{equation}

What does this curve look like?
Let me pass to Cartesian coordinates in the $xz$~plane,
\begin{equation}
\bar{R}^2 \= (2\,R_0)^2\,\cos\th \= x^2+z^2 \und 
\cos\th \= \sfrac{z}{\sqrt{x^2+z^2}} \ ,
\end{equation}
which yields the sextic curve (cubic in squares)
\begin{equation}
(x^2+z^2)^3 \= (2\,R_0)^4\,z^2 \qquad\textrm{with}\qquad
R_0^3 \= \sfrac{15}{32\pi}\,V \ .
\end{equation}
The parameter $R_0$ only takes care of the physical dimensions
and determines the overall size of the solid.
In dimensionless coordinates it may be put to unity,
which fixes the vertical diameter to be equal to~2 and allows
for a comparison of my optimal curve with the unit circle,
\begin{equation}
r(z) \= 2\,|z|^{1/3} 
\qquad\textrm{versus}\qquad
r(z)\= 2\,|z|^{1/2} 
\qquad\textrm{for}\quad r(z)^2\=x^2+z^2\ ,
\end{equation}
with $-z\in[0,2]$ and $r(z)\in[0,2]$.
Since $|z|^{1/3}\ge|z|^{1/2} $ in the interval of question, 
my curve lies entirely outside the reference circle, touching it only
twice on the $z$~axis.
(Note that $R_0\neq r_b$ so the corresponding volumes differ.)
Other than the sphere, my curve has a critical point: 
Due to $z\sim x^3$ near the origin, the curvature vanishes there.
Clearly, the vertical extension of~$\bar{B}$ is $\Delta z=2R_0$
while its width is easily computed to be
\begin{equation} 
\Delta x \= 2\,\root 4 \of {\sfrac{4}{27}}\;2R_0 \ \approx\ 2.48161\,R_0
\qquad\textrm{at}\qquad
z_0 \= -\root 4 \of {\sfrac{1}{27}}\;2R_0 \ \approx\ -0.87738\,R_0 \ .
\end{equation}
The shape of my optimal body~$\bar{B}$ vaguely resembles an apple,
with the flatter side up.

My final goal is to calculate the maximal possible weight $G_{\textrm{max}}$, or 
\begin{equation}
G[\bar R] \= 2\pi\,\gamma\,\sfrac{M}{V}\,2R_0 \int_0^1\!\diff\cos\th\ \bigl( \cos\th \bigr)^{3/2}
\= 2\pi\,\gamma\,\sfrac{M}{V}\,\root 3 \of {\sfrac{15\,V}{4\pi}}\;\sfrac25
\= \bigl( \sfrac{4\pi\sqrt{3}}{5} \bigr)^{2/3} \gamma\,M\,V^{-2/3}\ .
\end{equation}
Comparing with the spherical shape,
\begin{equation} \label{maximum}
\frac{G[\bar R]}{G_b} \= 3\cdot 5^{-2/3} \= \sfrac35\root 3 \of 5 \ \approx\ 1.02599 \ .
\end{equation}
I conclude that by homogeneous reshaping it is possible to increase the surface gravity of a spherical ball
by at most $2.6\%$ !

\newpage{}

\begin{figure}[!ht]
\begin{minipage}[t]{0.5\textwidth}
\begin{picture}(140,180)(60,165)
\put(0,0){\includegraphics[width=12cm,trim=0 0 0 60,clip]{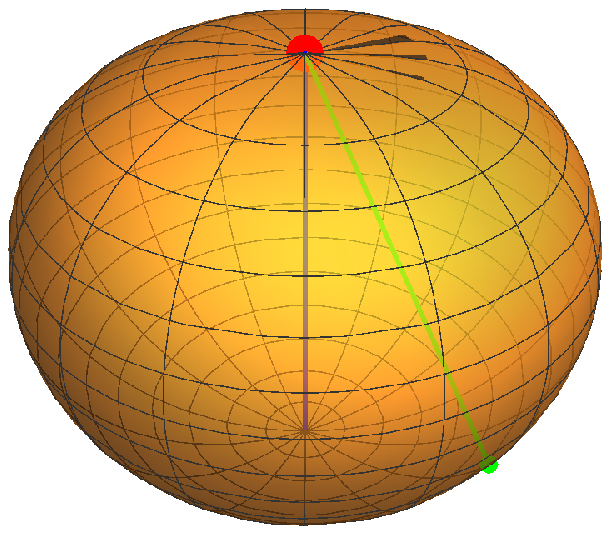}}
\end{picture}
\end{minipage}
\begin{minipage}[t]{0.5\textwidth}
\begin{picture}(140,180)(-15,20)
\put(0,0){\includegraphics[width=8cm]{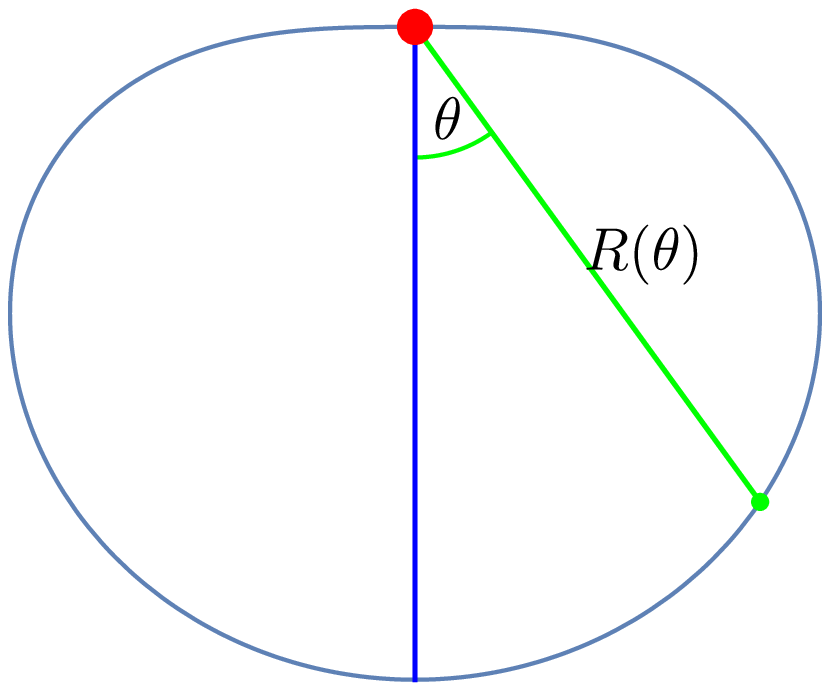}}
\end{picture}
\end{minipage}
\vspace{0.3cm}
\caption{ Optimal asteroid surface $\partial \bar{B}$}
\label{fig:4}
\end{figure}
%\vspace{0.5cm}

\section{Other shapes}

\noindent
Since the cylinder shape is already superior to the spherical one for maximizing surface gravity, 
it is interesting to explore a few other more or less regular bodies, 
to see how close they can get to the optimal value of $\sfrac35\root 3\of 5\approx1.02599$. 
Let me discuss three cases which are fairly easy to parametrize in the cylindrical coordinates chosen.

\begin{figure}[!ht]
\begin{minipage}[t]{0.5\textwidth}
\begin{picture}(140,140)(0,0)
\put(0,0){\includegraphics[width=7cm,trim=0 0 0 0,clip]{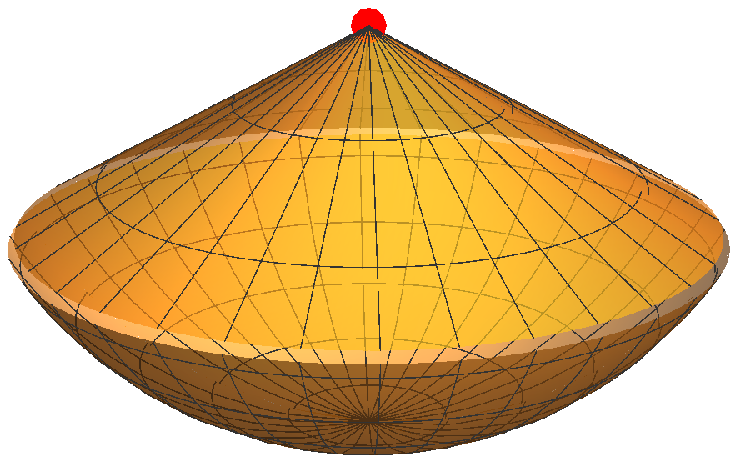}}
\end{picture}
\end{minipage}
\begin{minipage}[t]{0.5\textwidth}
\begin{picture}(140,140)(0,0)
\put(0,0){\includegraphics[width=8cm]{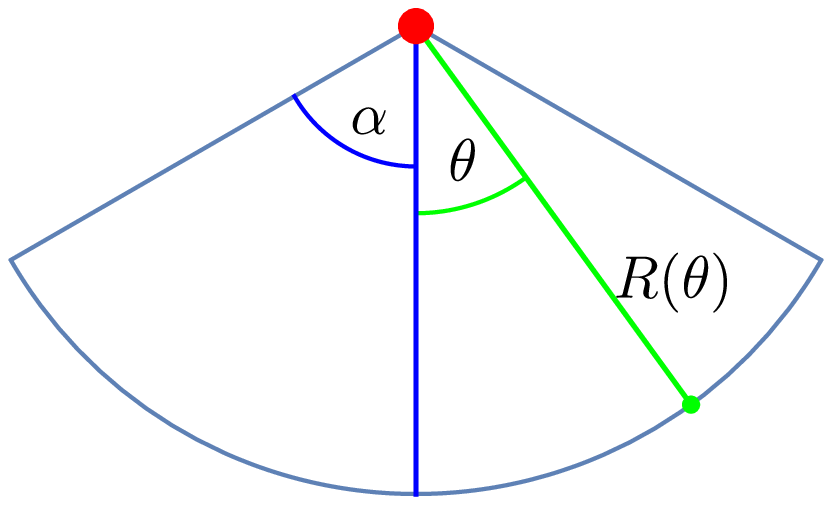}}
\end{picture}
\end{minipage}
%\vspace{0.5cm}
\caption{ Conical segment of a spherical ball}
\label{fig:5}
\end{figure}
\vspace{0.5cm}
First, I consider a conical segment of a spherical ball centered in the origin, 
with opening angle $2\al<\pi$ and radius $r_c$, see~Fig.~5. 
Here, one simply has
\begin{equation}
0 \le \th \le \al \und R_c(\th) \= r_c\ ,
\end{equation}
thus the surface gravity~(\ref{functional}) reduces to
\begin{equation}
G_c\= 2\pi\,\gamma\,\sfrac{M}{V} \int_{\cos\al}^1\!\diff\cos\th \int_0^{r_c}\!\diff{r}\;\cos\th 
\= 2\pi\,\gamma\,\sfrac{M}{V}\,r_c\,\sfrac12 (1-\cos^2\al) \ .
\end{equation}
\newpage{}
\begin{figure}[!ht]
\centerline{%\lower12pt\hbox{
\includegraphics[width=8cm]{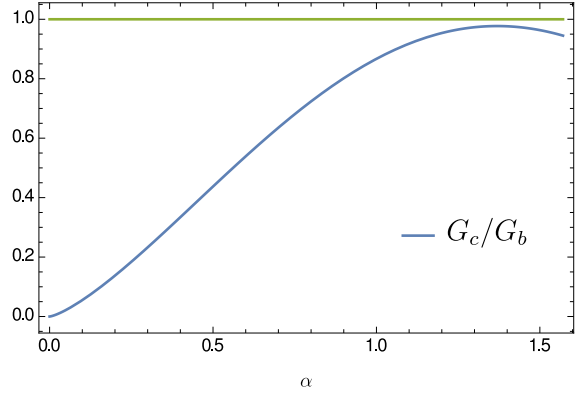}}%}
\caption{ Surface gravity on the apex of a conical segment of a spherical ball}
\label{fig:6}
\end{figure}
\vspace{0.5cm}
Since at the same time,
\begin{equation}
M\= 2\pi\,\sfrac{M}{V} \int_{\cos\al}^1\!\diff\cos\th \int_0^{r_c}\! r^2\diff{r}
\= 2\pi\,\sfrac{M}{V}\,\sfrac13 r_c^3\, (1-\cos\al) \ ,
\end{equation}
one gets
\begin{equation}
G_c \= \bigl( \sfrac{\sqrt{3}\pi}{\sqrt{2}} \bigr)^{2/3} \gamma\,M\,V^{-2/3}\,
(1-\cos^2\al)\,(1-\cos\al)^{-1/3}\ ,
\end{equation}
leading to the curve in Fig.~6,
\begin{equation}
\frac{G_c}{G_b} \= \sfrac34 \root 3\of 2\,(1-\cos^2\al)\,(1-\cos\al)^{-1/3}\ .
\end{equation} 
The best opening angle occurs at an angle of about $78.5^\circ$,
\begin{equation}
\cos\al \= \sfrac15 \ \approx\ 1.36944 \qquad\Rightarrow\qquad
\frac{G_c}{G_b}\bigg|_{\textrm{max}} \= 2^{2/3}\cdot 9\cdot 5^{-5/3}
\ \approx\ 0.97719\ .
\end{equation}
Clearly, the spherical ball beats any cone.
The value $\al=\sfrac{\pi}{2}$ describes a semi-ball, which yields
\begin{equation}
\frac{G_c}{G_b}\bigg|_{\al=\pi/2} \= 2^{-8/3}\cdot 3\ \approx\ 0.94494\ .
\end{equation}

\begin{figure}[!ht]
\begin{minipage}[t]{0.5\textwidth}
\begin{picture}(140,240)(-20,-10)
\put(0,0){\includegraphics[width=7cm,trim=0 0 0 0,clip]{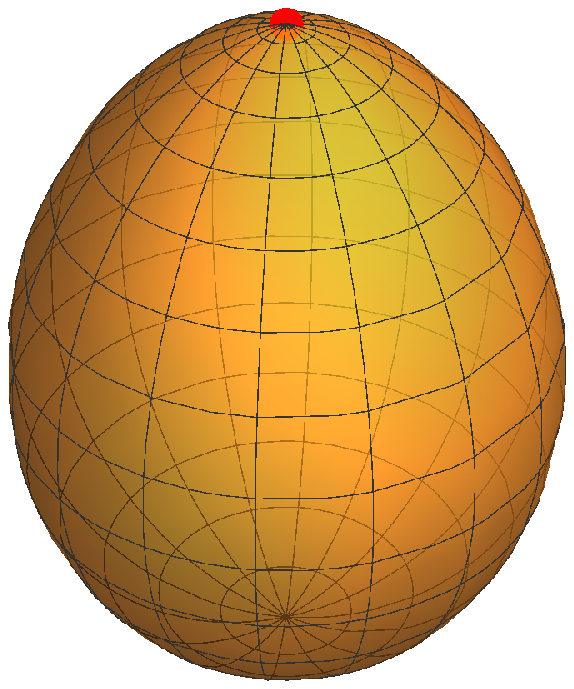}}
\end{picture}
\end{minipage}
\begin{minipage}[t]{0.5\textwidth}
\begin{picture}(140,240)(-20,0)
\put(0,0){\includegraphics[width=7cm]{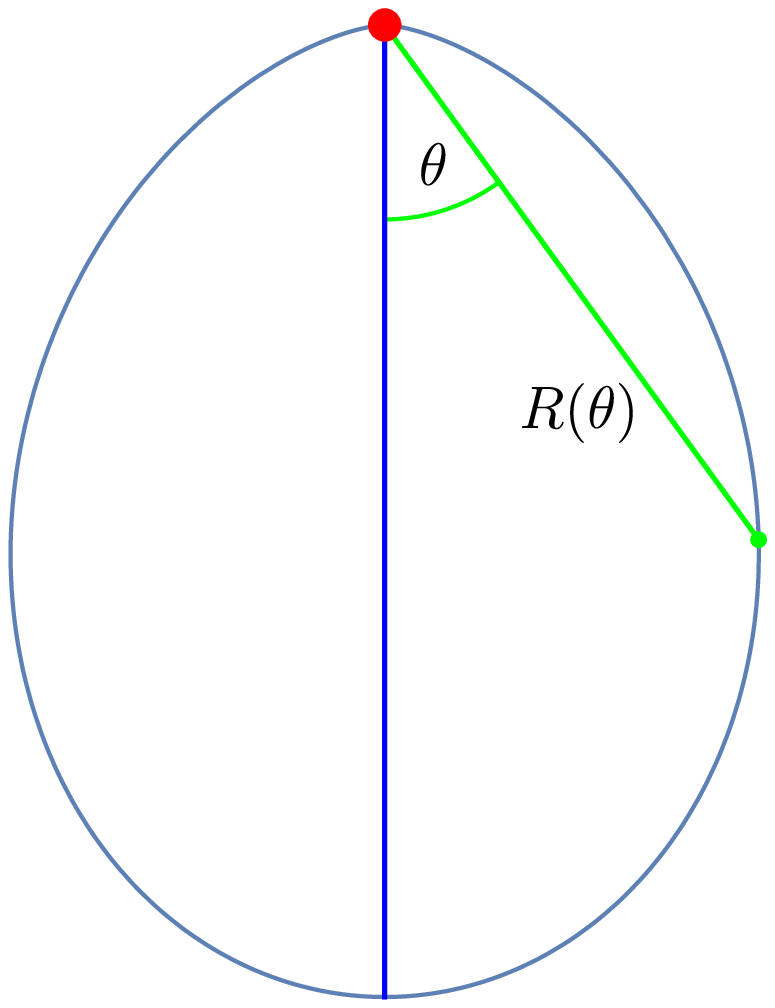}}
\end{picture}
\end{minipage}
%\vspace{0.5cm}
\caption{ Shape for radius function $R(\th)\sim\cos^2\th$}
\label{fig:7}
\end{figure}
\vspace{0.5cm}
Second, let me try out the radius function $R(\th)$ 
being an arbitrary power $n$ of~$\cos\th$,
\begin{equation}
R_n(\th) \= 2\,r_n\,(\cos\th)^n \qquad\textrm{with}\quad n>0 \ ,
\end{equation}
displayed in Fig.~7 for $n{=}2$. 
This produces
\begin{equation}
G_n \= 2\pi\,\gamma\,\sfrac{M}{V}\,2r_n\int_0^1\!\diff\cos\th\ (\cos\th)^{n+1}
\= 2\pi\,\gamma\,\sfrac{M}{V}\,2r_n\,\sfrac1{n+2}\ .
\end{equation}
The special value of $n{=}1$ yields a spherical ball, which separates 
squashed forms ($n{<}1$) from elongates ones ($n{>}1$). With
\begin{equation}
M\= \sfrac{2\pi}{3}\,\sfrac{M}{V} \int_0^1\!\diff\cos\th\ \bigl(2r_n\,(\cos\th)^{n}\bigr)^3
\= \sfrac{2\pi}{3}\,\sfrac{M}{V}\,(2r_n)^3\,\sfrac{1}{3n+1}
\end{equation}
I can eliminate $r_n$ and find
\begin{equation}
\frac{G_n}{G_b} \= 3\,\bigl( \sfrac14(3n+1)\bigr)^{1/3} \big/ (n+2)\ ,
\end{equation}
which is shown in Fig.~8.
This is indeed maximized for
\begin{equation}
n=\sfrac12 \qquad\Rightarrow\qquad
\sfrac{G_{1/2}}{G_b} \= 3\cdot 5^{-2/3}\ ,
\end{equation}
as was already found in (\ref{solution}) and (\ref{maximum}). 
It exceeds unity in the interval $0.17424\lesssim n < 1$.
\vspace{0.5cm}
\begin{figure}[!ht]
\centerline{%\lower12pt\hbox{
\includegraphics[width=8cm]{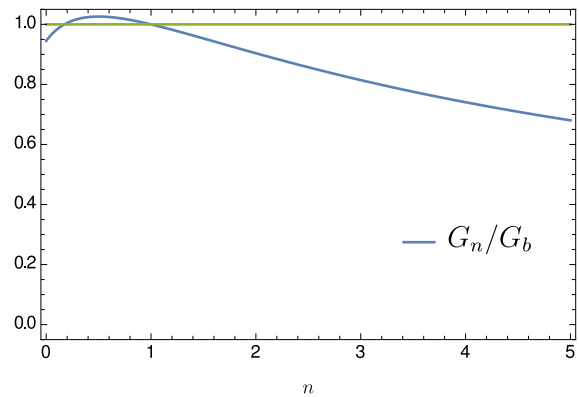}}%}
\caption{ Surface gravity on a body with radius function $R(\th)\sim\cos^n\th$}
\label{fig:8}
\end{figure}

\newpage{}

%\vspace{0.5cm}
\begin{figure}[!ht]
\begin{minipage}[t]{0.5\textwidth}
\begin{picture}(140,150)(-20,10)
\put(0,0){\includegraphics[width=7cm,trim=0 0 0 0,clip]{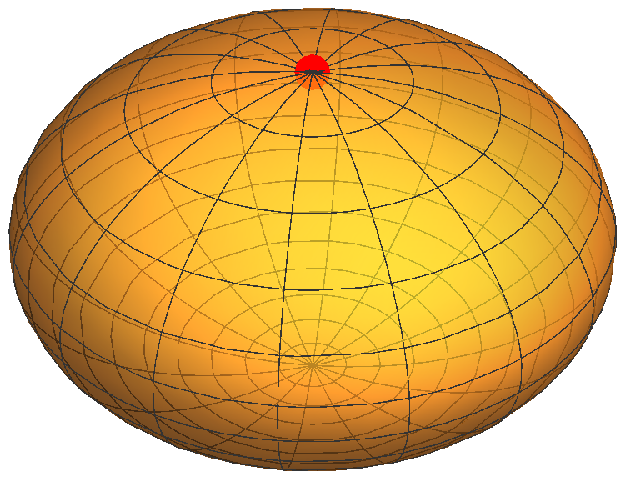}}
\end{picture}
\end{minipage}
\begin{minipage}[t]{0.5\textwidth}
\begin{picture}(140,150)(-20,0)
\put(0,0){\includegraphics[width=7cm]{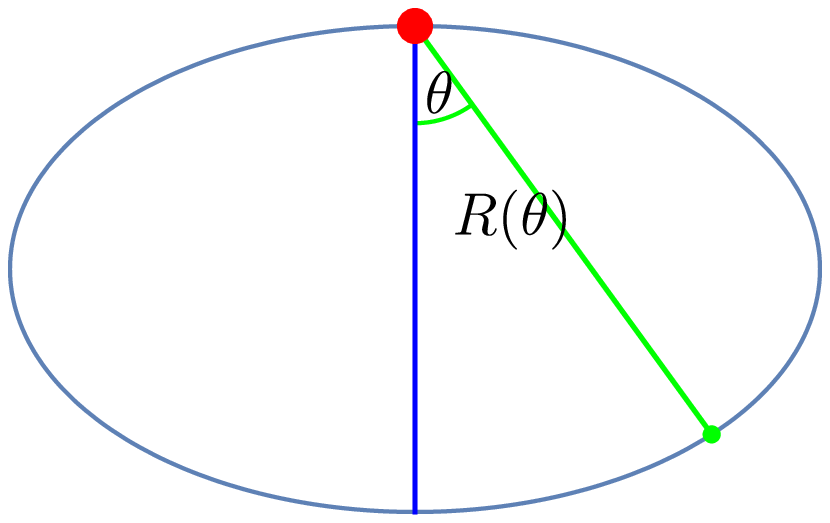}}
\end{picture}
\end{minipage}
%\vspace{0.5cm}
\caption{ Oblate ellipsoid with eccentricity $\eps=0.8$}
\label{fig:9}
\end{figure}
\vspace{0.5cm}
Third, I look at an oblate ellipsoid of revolution 
with minor semi-axis length $r_e$ and eccentricity $\eps$,
see Fig.~9. In this case,
\begin{equation}
R_e(\th) \= \frac{2\,r_e\,\cos\th}{1-\eps^2\sin^2\th}
\= \frac{2\,r_e\,\cos\th}{1{-}\eps^2+\eps^2\cos^2\th}
\qquad\textrm{with}\quad \eps\in[0,1)\ ,
\end{equation}
which includes the sphere for $\eps{=}0$.
(The prolate case corresponds to imaginary~$\eps$.)
The surface gravity and mass integrals then become
\begin{equation}
G_e \= 2\pi\,\gamma\,\sfrac{M}{V}\,2r_e\int_0^1\!\diff{y}\
\frac{y^2}{1{-}\eps^2+\eps^2 y^2}
\= 2\pi\,\gamma\,\sfrac{M}{V}\,2r_e\,\frac1{\eps^2}
\Bigl( 1 - \sqrt{\sfrac{1-\eps^2}{\eps^2}}\arctan\sqrt{\sfrac{\eps^2}{1-\eps^2}} \Bigr)\ ,
\end{equation}
\begin{equation}
M \= \sfrac{2\pi}{3}\,\sfrac{M}{V}\,(2r_e)^3\int_0^1\!\diff{y}\
\frac{y^3}{(1{-}\eps^2+\eps^2 y^2)^3}
\= \sfrac{2\pi}{3}\,\sfrac{M}{V}\,(2r_e)^3\,\frac{1}{4(1{-}\eps^2)}\ ,
\qquad\qquad\qquad\quad{}
\end{equation}
respectively. From this I conclude that
\begin{equation}
\frac{G_e}{G_b} \= 3\,\bigl(1-\eps^2\bigr)^{1/3}\,\frac1{\eps^2}
\Bigl( 1 - \sqrt{\sfrac{1-\eps^2}{\eps^2}}\arctan\sqrt{\sfrac{\eps^2}{1-\eps^2}} \Bigr)\ ,
\end{equation}
shown in Fig.~10.
This is larger than one for $\eps\lesssim0.85780$ and is maximized numerically at
\begin{equation}
\eps\ \approx\ 0.69446 \qquad\Rightarrow\qquad
\frac{G_e}{G_b}\bigg|_{\textrm{max}} \ \approx\ 1.02204 \ .
\end{equation}
Hence, I can come to within less than $0.4\%$ of the optimal surface gravity by 
engineering an appropriate ellipsoid.

\newpage{}

\begin{figure}[!ht]
\centerline{%\lower12pt\hbox{
\includegraphics[width=8cm]{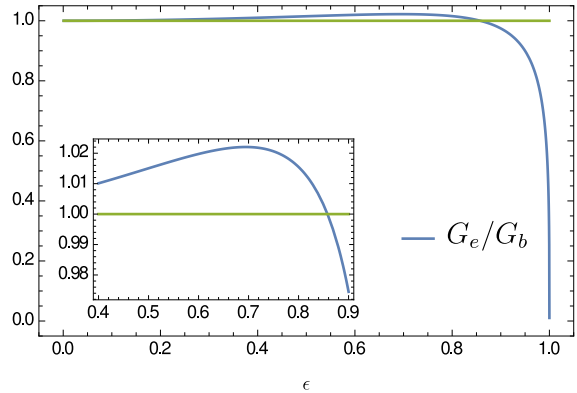}}%}
\caption{ Surface gravity on an ellipsoid with eccentricity~$\eps$}
\label{fig:10}
\end{figure}
%\vspace{0.5cm}

\section{Conclusions}

\noindent
The main result of this short paper is a universal sixth-order planar curve,
\begin{equation}
{\cal C}_{\textrm{Gh}}: \quad
(x^2+z^2)^3-(4\,z)^2\=0 \qquad\Leftrightarrow\qquad r(\th)\=2\sqrt{\cos\th}\ ,
\end{equation}
which characterizes the shape of the homogeneous body admitting the maximal possible 
surface gravity in a given point, for unit mass density and volume.
It is amusing to speculate about its use for asteroid engineering in an advanced civilization
or our own future. This curve seems not yet to have occurred in the literature,
and so I choose to name it ``Gharemani curve'' after my deceased friend who 
initiated the whole enterprise.

The maximally achievable weight on bodies of various shapes is listed in the following table.
It occurs at the intersection of the rotational symmetry axis with the body's surface
and is normalized to the value on the spherical ball.
\begin{center}
\begin{tabular}{|c|ccccc|}
\hline
shape & cone & ball & cylinder & ellipse & Gharemani \\
\hline
maximum of $G/G_b$ & 0.97719 & 1.00000 & 1.00682 & 1.02204 & 1.02599 \\
\hline
\end{tabular}
\end{center}

It can pay off to get inspired by the curiosity of your non-scientist friends. 
The result is a lot of fun and may even lead to new science!

\begin{figure}[!ht]
\centerline{%\lower12pt\hbox{
\includegraphics[width=6cm]{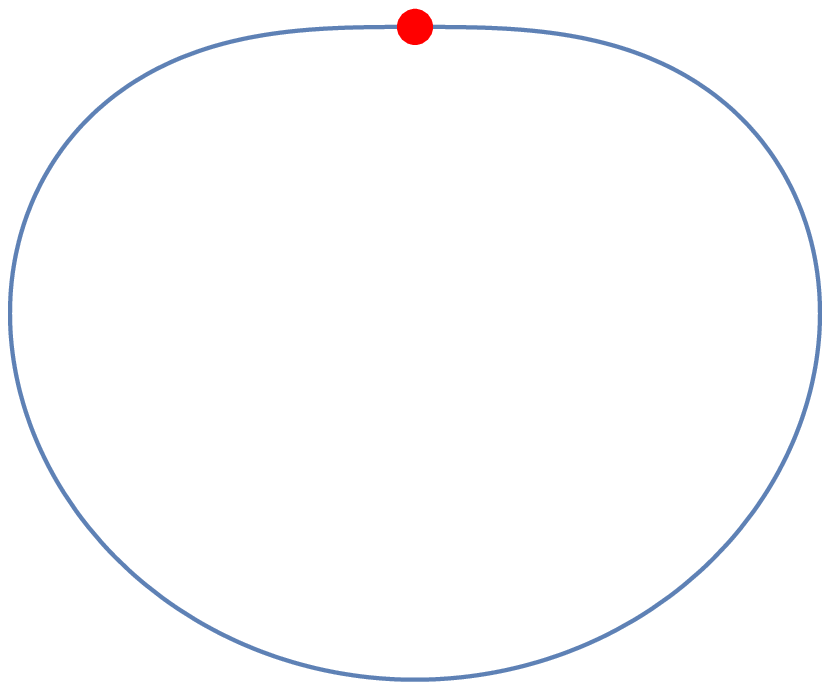}}%}
%\caption{ The Gharemani curve}
%\label{fig:11}
\end{figure}

%\bigskip
\newpage

\section*{Acknowledgments}
\noindent
I thank Michael Flohr for help with Mathematica and the integral~(\ref{hardintegral}).

\bigskip

\small{

}

\end{document}